\documentclass[universe,article,accept,pdftex,moreauthors]{Definitions/mdpi} 

\usepackage{verbatim}
\usepackage{aas_macros}

\firstpage{1} 
\pubvolume{1}
\issuenum{1}
\articlenumber{0}
\pubyear{2024}
\copyrightyear{2024}
\externaleditor{Academic Editor: }
\datereceived{27 July 2024 } 
\daterevised{20 August 2024 } 
\dateaccepted{22 August 2024 } 
\datepublished{ } 

\hreflink{https://doi.org/}

\Title{Characterisation of the Atmosphere in Very High Energy
 Gamma-Astronomy
for Imaging Atmospheric Cherenkov Telescopes}
\TitleCitation{Characterisation of the Atmosphere in Very High Energy Gamma-Astronomy
for Imaging Atmospheric Cherenkov Telescopes}

\Author{Dijana Dominis Prester 
 $^{1,}$*\orcidA{}, Jan Ebr $^{2}$\orcidE{}, Markus Gaug $^{3}$\orcidB{}, Alexander Hahn $^{4}$\orcidC{}, Ana Babić $^{5}$\orcidJ{}, Ji\v{r}\'{i} Eli\'{a}\v{s}ek 
 $^{2}$\orcidH{},\linebreak  Petr Jane\v{c}ek $^{2}$\orcidF{}, Sergey Karpov $^{2}$\orcidI{}, Marta Kolarek $^{6}$\orcidK{}, Marina Manganaro $^{1,}$*\orcidG{} and Razmik Mirzoyan $^{4}$\orcidD{}
}

\AuthorNames{Dijana Dominis Prester, Jan Ebr, Markus Gaug, Alexandar Hahn, Ana Babić, Ji\v{r}\'{i} Eli\'{a}\v{s}ek, Petr Jane\v{c}ek, Sergey Karpov, Marta Kolarek, Marina Manganaro and Razmik Mirzoyan}

\AuthorCitation{Dominis Prester, D.;  Ebr, J.; Gaug, M.; Hahn, A.; Babić, A.; Eli\'{a}\v{s}ek, J.; Jane\v{c}ek, P.; Karpov, P.;  Kolarek, M.; Manganaro, M.; et al.}

\address{
$^{1}$ \quad Faculty of Physics,  University of Rijeka, Radmile Matejčić 2, 51000 Rijeka, Croatia\\
$^{2}$ \quad FZU---Institute of Physics of the Czech Academy of Sciences, Na Slovance 1999/2, 18200 Prague, Czech Republic; ebr@fzu.cz (J.E.); eliasek@fzu.cz (J.E.); janecekp@fzu.cz (P.J.); karpov@fzu.cz (S.K.) \\
$^{3}$ \quad Departament de F\'isica and CERES-IEEC, Universitat Aut\`onoma de Barcelona, 08193 Bellaterra, Spain; markus.gaug@uab.cat\\
$^{4}$ \quad Max Planck Institut fuer Physik, 80805 Munich, Germany; ahahn@mpp.mpg.de (A.H.); razmik@mpp.mpg.de (R.M.)\\
$^{5}$ \quad Faculty of Electrical Engineering and Computing University of Zagreb, Unska 3, 10000 Zagreb, Croatia; ana.babic@fer.unirzg.hr\\
$^{6}$ \quad Department of Medical Physics, The University Hospital Centre Zagreb, Kišpatićeva 12, 10000 Zagreb, Croatia; marta.kolarek@kbc-zagreb.hr\\}

\corres{Correspondence: dijana@uniri.hr (D.D.P.); marina.manganaro@uniri.hr (M.M.)}

\abstract{Ground-based observations of Very High Energy (VHE) gamma rays from extreme astrophysical sources are significantly influenced by atmospheric conditions. This is due to the atmosphere being an integral part of the detector when utilizing Imaging Atmospheric Cherenkov Telescopes (IACTs). Clouds and dust particles diminish atmospheric transmission of Cherenkov light, thereby impacting the reconstruction of the air showers and consequently the reconstructed gamma-ray spectra.
Precise measurements of atmospheric transmission above Cherenkov observatories play a pivotal role in the accuracy of the analysed data, among which the corrections of the reconstructed energies and fluxes of incoming gamma rays, and in establishing observation strategies for different types of gamma-ray emitting sources. The Major Atmospheric Gamma Imaging Cherenkov (MAGIC) telescopes and the Cherenkov Telescope Array Observatory (CTAO), both located on the Observatorio del Roque de los Muchachos (ORM), La Palma, Canary Islands, use different sets of auxiliary instruments for real-time characterisation of the atmosphere.
In this paper, historical data taken by MAGIC LIDAR (LIght Detection And Ranging) and CTAO FRAM (F/Photometric Robotic Telescope) are presented. From the atmospheric aerosol transmission profiles measured by the MAGIC LIDAR and CTAO FRAM aerosol optical depth maps, we obtain the characterisation of the clouds above the ORM at La Palma needed for data correction and optimal observation scheduling.} 


\begin{document}

\section{Introduction} 

Imaging Atmospheric Cherenkov Telescopes (IACTs) observe astrophysical sources, which emit Very High Energy (VHE) gamma rays in the energy range from GeV up to TeV~\citep{hinton2009}. By observing Cherenkov radiation produced by air showers, which are themselves caused by gamma-ray photons from space impinging on the Earth's atmosphere, the energy and direction of the incoming gamma rays are reconstructed. 
The atmosphere is hence an intrinsic part of the detection method of IACTs. The characterisation and monitoring of the atmospheric conditions and their effects on the Cherenkov radiation travelling from its respective emission height to ground play a crucial role in the proper working of the IACT method and consequently the understanding of the physics of extreme astrophysical sources and phenomena studied by IACTs, such as Active Galactic Nuclei (AGNs), Gamma-Ray Bursts (GRBs), pulsars, supernova remnants, microquasars, interacting stellar binary systems, cosmic rays, and in studies of dark matter and Lorenz Invariance Violation (LIV)~\citep{ScienceWithCTA}.
IACT observations of these different types of astrophysical sources and phenomena, at different distances from Earth, compete for the same observation time windows.
However, impaired atmospheric conditions can lead to strongly energy-dependent effects on sensitivity~\citep{Sobczynska:2015}.  
Therefore, complex observational strategies have been developed for both MAGIC and CTAO, with the aim of optimising the use of the total observing time and scientific output of various observing proposals. This, in turn, has increased the complexity of observation scheduling strategies and their implementations.  The zenith angle of the source, moonlight, and atmospheric parameters strongly affect the sensitivity of the telescopes~\citep{MAGIC_performance_II_2016,MAGICMoon,schmuckermaier_correcting_2023}, unequally for different energy ranges.

Astrophysical sources are often observed by IACTs using a Target of Opportunity (ToO) observational strategy, especially in cases of short-term variability or when the quiet-state flux is not detectable by the IACT. In such cases, observations are triggered by alerts from other IACTs, or more often, communicated by ground-based or space telescopes observing at other energies of the electromagnetic spectrum~\citep{2024Univ...10...80M}, or by detection of other messengers like neutrinos~\citep{2018Sci...361.1378I}, following a so-called “multi-messenger” observational strategy~\citep{Multimessenger-COST-2022}.

Active Galactic Nuclei (AGNs) are found at a huge range of distances from Earth, where close-by AGNs are visible up to TeV energies and beyond~\citep{Mrk421-MAGIC:2020}, whereas the light from distant AGNs suffers from strong extragalactic background light absorption~\citep{Dominguez:2011} and can be detected only at lower energies, sometimes limited down to $<$100~GeV. Moreover, AGNs show both long-term and short-term time variability ~\citep{1997ARA&A..35..445U}. In cases of long-term variability, selected AGNs are monitored over a longer period of time, years to decades in some cases, with so-called ``snapshots'' taken over regular time intervals. VHE gamma-ray observations are often coordinated with instruments observing in other parts of the electromagnetic spectrum, following a so-called multiwavelength (MWL) observing strategy. 

In searches for signatures of dark matter or the detection of pulsars in the VHE energy range, many hours of observations need to be collected over longer periods of time. For these studies, sensitivity at lower energies is crucial, meaning that observing conditions with moderate atmospheric transmission (i.e., presence of clouds) are not suitable, and many valuable observing hours could be potentially wasted. Instead, low-redshift or flaring AGN observations can be accommodated.

In order to schedule all these different observations while optimising the total observational time of an IACT, real-time measurements of the atmospheric aerosol transmission profile, a detailed characterisation of the clouds above the observatory, and an understanding of the impact of the clouds on the development of air showers and reconstructed energies~\citep{Pecimotika:2023} are important. In addition, for the proper reconstruction of the gamma-ray energy spectrum of the observed astrophysical source, corrections calculated using transmission profile measurements should be implemented at the level of the data analysis~\citep{Sobczynska:2020,Zywucka:2024}.

Aerosols in the atmosphere have a substantial impact on climate, weather, and air quality. Accurate parameters are required to evaluate the distribution of aerosols and their interaction with electromagnetic radiation.  
The atmospheric extinction coefficient is the sum of scattering and absorption coefficients, both by particles and gases. It is a measure of the alteration of radiant energy as it passes through the atmosphere and depends on the aerosol mass concentration, size distribution, shape, and refractive index. The aerosol optical depth (AOD) is defined as the integral of the aerosol extinction coefficient over the entire line of sight up to the top of the atmosphere. Then, $\exp(-\mathrm{AOD})$ measures the relative amount of light passing through all aerosols and clouds if emitted from a source at infinity. Note that the AOD depends on the wavelength of light. 
The vertical aerosol optical depth (VAOD) is defined as an AOD for vertically incident light and becomes independent of the measuring device's pointing direction if aerosols are stratified. 

In this work, atmospheric monitoring strategies and different instruments used for the characterisation of the atmosphere for IACTs are described, with an emphasis on the optical LIDAR (LIght Detection And Ranging) used within the MAGIC experiment and the FRAM (F/Photometric Robotic Telescope) used for the CTAO. Historical data from both the MAGIC LIDAR and the CTAO FRAM are newly analysed. Based on data measured by the MAGIC LIDAR, aerosol/cloud transmission profiles are generated, whereas aerosol optical depth maps are calculated from the CTAO FRAM data. From comparing the results of both instruments, we aim to obtain reliable, high-cadence online monitoring and a detailed characterisation of aerosols and clouds above the CTAO, which can be used for both optimal observation scheduling and data correction.

\section{Atmospheric Monitoring Strategies and Instruments for Imaging Atmospheric Cherenkov Telescopes}

\begin{figure}[H]
\includegraphics[width=0.9\linewidth]{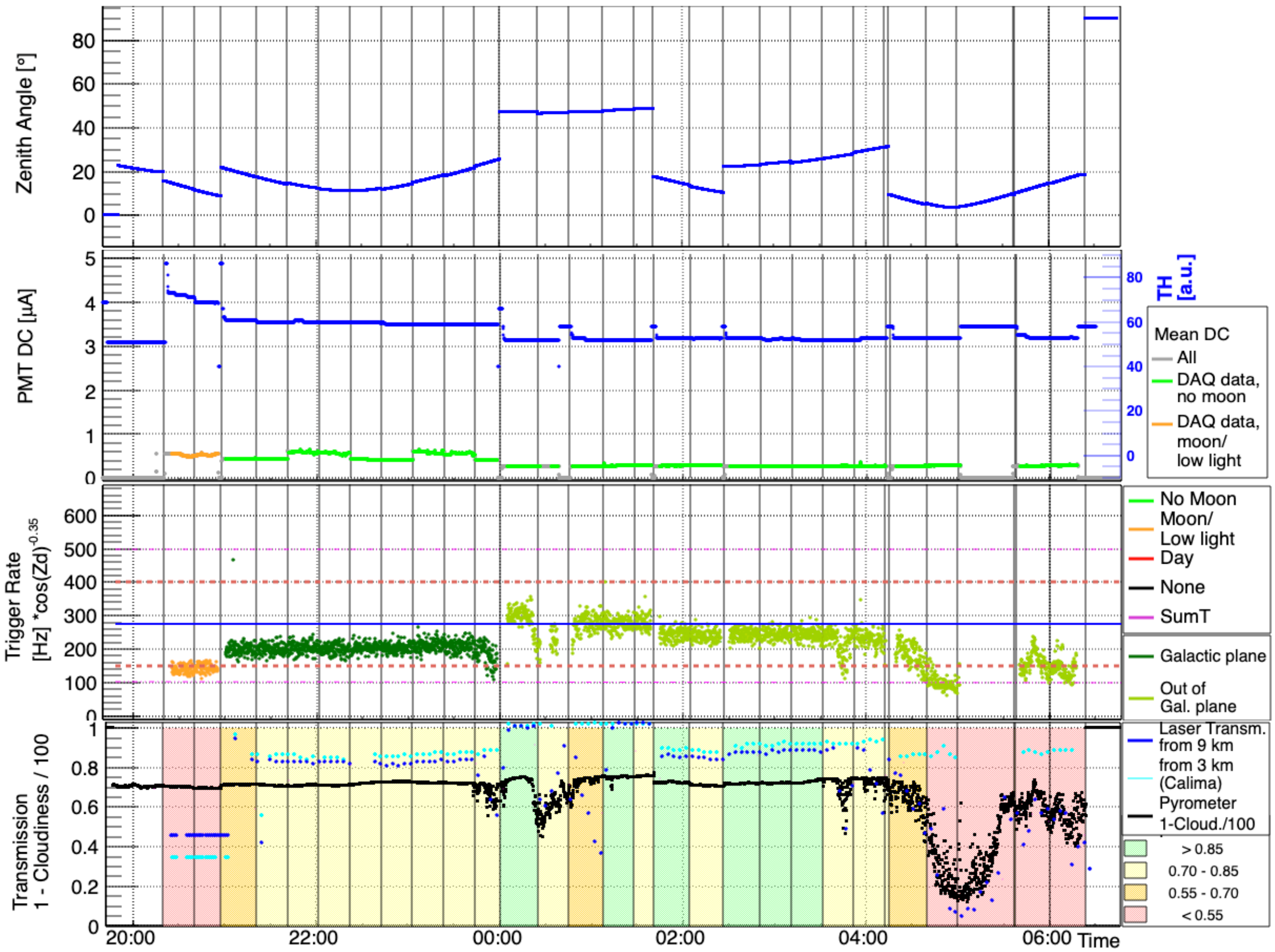}
\caption{Overview plot used in data quality checks for a single observing night at the MAGIC telescope site. Four panels (enumerated from top to bottom) allow for an overview of the most important parameters, which describe the data-taking conditions and/or impact on the data quality. Panel 1: Zenith angle of the telescope pointing, with vertical lines marking the wobble position or source changes. 
Panel 2: Mean direct current (DC; green and red) and the median discriminator threshold (TH; blue) for the camera PMTs. The colour of the DC plot roughly captures the dark time (green) and moon time (orange) observations, with dark time observations further separated into extragalactic (dark green) and galactic (light green).  Panel 3: Coincidence trigger rates between the two MAGIC telescopes. Panel 4: Atmospheric aerosol transmission, based on measurements by the LIDAR and the pyrometer. Two transmission measures are given based on the LIDAR profiles, from 3 km and from 9 km above the site, quantifying the dust intrusions (calima) and cloud effects, respectively. Colour shading gives a rough guidance whether the data may be used without corrections, with nominal or increased energy threshold (green, light yellow), may be corrected for atmosphere effects (dark yellow), or cannot be used at all due to poor quality (red).
\label{MAGIC_superplot}}
\end{figure}

Since atmospheric conditions above an IACT may change over time scales of minutes, particularly due to the passage of cirrus clouds above the telescopes~\citep{fruck_characterizing_2022,schmuckermaier_correcting_2023}, atmospheric transmission must be monitored on similar time scales. Likewise, active-sensing instruments, such as LIDARs, may impact the science observations of IACTs, e.g., through spurious triggers from a laser pulse. Several previous studies~\citep{Holch:2022,schmuckermaier_correcting_2023,Pecimotika:2023} have shown that the height information of atmospheric disturbances must be nevertheless taken into account for a successful reconstruction of air showers and hence the properties of the initial gamma ray. These considerations have led to the use of elastic LIDARs~\citep{Bregeon:2016,fruck_characterizing_2022}, pyrometers~\citep{Will:2017,connolly:phd}, or radiometers~\citep{Hahn2015} and the use of data rates~\citep{Hahn2014,Will:2017} for the current (second) generation of IACTs.
Figure~\ref{MAGIC_superplot} shows how atmospheric information from the different instruments and methods can be combined.

Lessons learnt from that rather large variety of monitoring instruments have led to the decision to use a combination of Raman LIDARs~\citep{Ballester:2019,Vasileiadis:2020,zivec:2022} and stellar photometers (the so-called “FRAM”~\citep{2021AJ....162....6E}) for the CTAO. 
One of the key elements of this strategy is to assess the main characteristics of each site well in advance and then construct tailored devices and methods to monitor each atmospheric parameter. Contrary to previous IACTs, aerosol and cloud monitoring devices must cover a wide field of view of more than 10$^\circ$$\times$10$^\circ$. 

Requirements for such instruments include the ability to measure the aerosol transmission profile along any line of sight, down to elevations of only 25$^\circ$, in at least two wavelength bands that span the observed Cherenkov light spectrum (between 300~nm and 600~nm) to an absolute accuracy of better than 0.03~rms and an altitude and distance resolution of better than 150~m rms. At the same time, the \AA ngstr\"om exponent should be determined with an absolute accuracy of better than $\pm$0.3~rms. Measurements fulfilling these characteristics should be carried out in less than a minute.  Given the high altitudes of cirrus clouds observed at the northern site of the CTAO, the Observatorio del Roque de los Muchachos (ORM), of up to 20~km a.s.l.~\citep{Gaug:2022}, these requirements entail, on one side, powerful lasers, in order to be able to determine distant cirrus clouds with sufficient signal-to-noise ratio, and on the other hand, systems with a full overlap at short distances, and hence a rather large pinhole field of view. Ideas to use the very same IACT mirrors, together with a low-power laser and a modified central camera pixel, actually go in the opposite direction and are probably incompatible with the foreseen operation of the CTAO. The CTAO opted, instead, for systems specifically designed for the mentioned requirements.

\begin{figure}[H]
\includegraphics[width=0.9\textwidth]{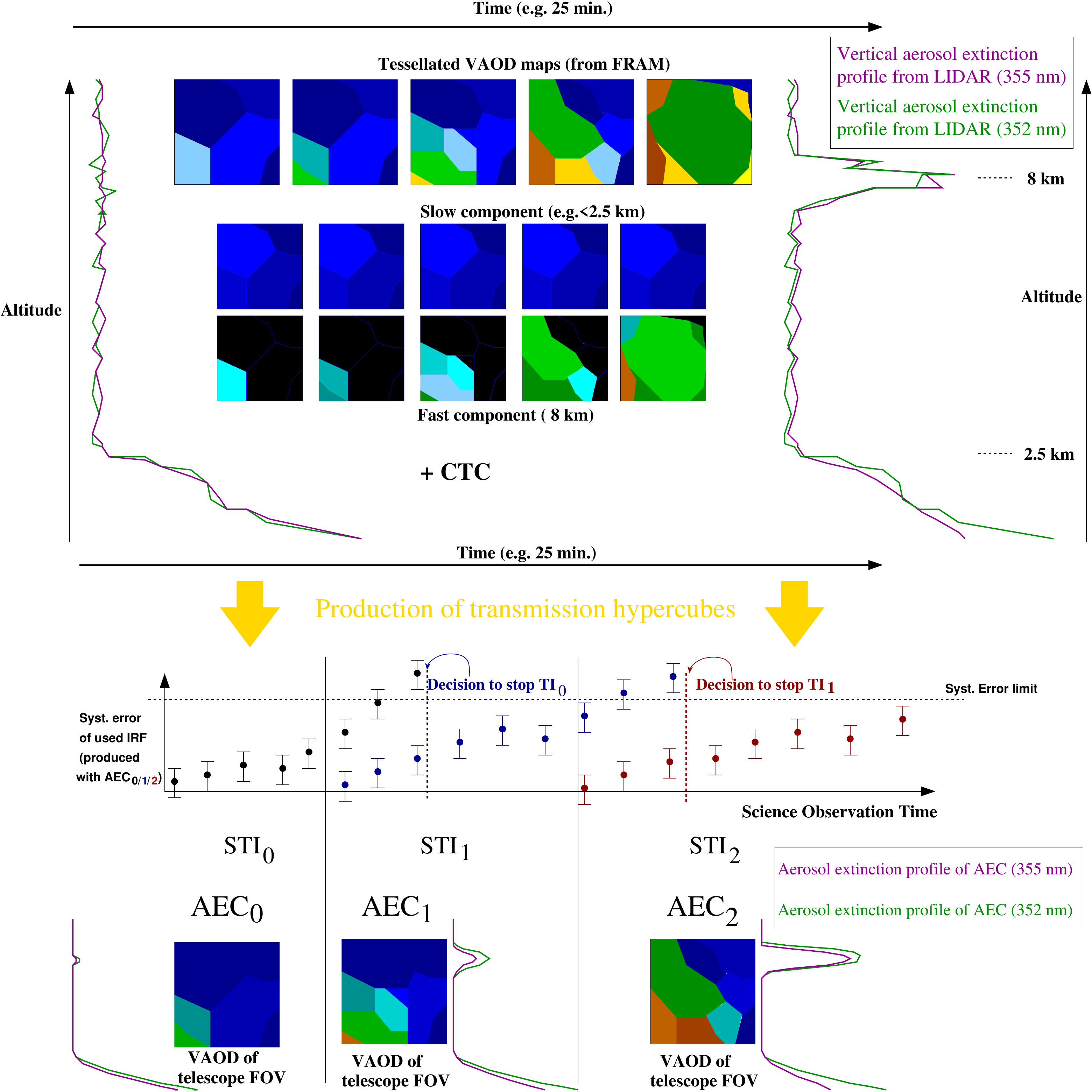}
\caption{\label{fig:atmocorrections}Scheme for the determination of new Stable Time Intervals (STIs) and new Monte Carlo (MC) simulated instrument response functions (IRFs): on the top, two possible aerosol extinction profiles at two wavelengths obtained from LIDAR are sketched, with nested VAOD maps obtained from FRAM during a typical CTAO observation block of 25~min. In this particular case, a typical aerosol ground layer of a clear night that ranges up to 2.5~km above ground evolves slowly towards a higher \AA ngstr\"om exponent (smaller average particulate sizes) at a similar vertical profile, whereas a cloud at a typical height of about 10~km moves into the FRAM field of view. The joint analysis of FRAM and LIDAR (and possibly CTC) data divides the tessellated VAOD maps from FRAM first into a slow and a fast component and then produces the aerosol transmission hypercubes through interpolation, both in time and in altitude. The offline science data analysis calculates systematic errors introduced by the not perfectly matching aerosol conditions and the aerosol extinction cube used for the production of the given IRFs, and decides to start new STIs as the systematic error exceeds a predefined limit. Each STI is then analysed with its (simplified) AEC used for the production of a tailored IRF. See text for further details. }
\end{figure}

Figure~\ref{fig:atmocorrections} then highlights the procedure for obtaining average instrument response functions over a time interval within which the systematic error due to simplifications of the profile remains acceptable.

Measurements of the full profile are taken with a Raman LIDAR before and after an observation time slot. Between these, the wide-angle stellar photometer FRAM follows the field of view of the CTAO array(s) and acquires maps of aerosol optical depth (AOD), for which the sky is divided into appropriate tiles (e.g., using a Voronoi tessellation technique). 
The tiled AOD maps are then interpolated in time. Interlaced Raman LIDAR measurements may serve to divide the interpolated AODs into vertical bins and to recalibrate its integral. These “aerosol transmission hypercubes” (with altitude, wavelength, and time as additional dimensions) are then split into a slow component (due to molecular extinction and quasi-stable aerosol layers, like the boundary layer or sometimes even clouds) and a fast one, which changes throughout a science observation block (OB). Historical information from aerosol characterisation campaigns and later on the very same database of previous LIDAR and photometer observations, together with now-cast predictions of the molecular profile, helps to produce the transmission hypercubes and in the identification of their components.
The expected shifts in the CTAO performance from a given extinction hypercube are continuously confronted with those of currently used (or foreseen) simulated instrument response functions (IRFs), and a series of systematic errors are calculated and confronted with the CTAO requirements. In case one of these systematic errors exceeds the allowed threshold, a new “Stable Time Interval” (STI) is communicated to the CTAO analysis software, together with a suitable start time, and an averaged extinction hypercube, to be used for a new MC simulation of the atmosphere. The average extinction cube (AEC) contains the vertical profile in one dimension, the wavelength dependence in a second dimension, and the aerosol extinction coefficient as a result of both. The AEC should be quality-checked and confronted with alternative procedures, such as the Cherenkov transparency coefficient (CTC)~\citep{CTC:2019}.
The critical part of this procedure is a robust estimate of the systematic error. For this purpose, MC simulation studies are required to determine the impact of aerosols and clouds at different altitudes and cirrus clouds covering parts of the gamma-ray field of view on the angular and energy resolution and bias~\citep{Pecimotika:2023}. Studies of the typical morphology of clouds and their evolution time scales at the CTAO sites are needed as input for these simulations, a task presented in Section~\ref{sec:profiles}.

In addition, the future short-term scheduler of the CTAO will benefit from the information gathered by instruments monitoring the state of the atmosphere beyond the currently observed field of view and time interval: an all-sky camera with 2$\pi$ field of view and capable of performing stellar extinction analysis~\citep{Mandat:2017,Adam:2017} and a ceilometer that allows us to determine the altitude (profiles) of clouds observed towards scheduled future fields of view. 
We know that the scheduling strategies of current IACTs would have great potential for improvement, if atmospheric conditions were taken into account for the specific science case behind an observation. This applies particularly to sources with very soft spectra observable below about 100 GeV only, which require absolutely clear nights (so-called “photometric nights” in optical astronomy) in order to escape later offline data rejection. Similar arguments apply to precision-pointing observations. Whereas data from other targets at higher energies can be rather easily corrected, these low-energy events are lost since they do not even trigger the readout.

\subsection{MAGIC Optical LIDAR}

MAGIC is a stereoscopic system of two IACTs~\citep{MAGIC_performance_II_2016}. The telescopes are located at the Observatorio del Roque de los Muchachos (ORM), on the Canary Island of La Palma. Since 2013, an absolutely calibrated elastic LIDAR system has been used to characterise aerosol transmission above the telescopes~\citep{fruck_characterizing_2022}. The LIDAR currently consists of a passively Q-switched, frequency-doubled Nd:YAG laser with a $25\,\upmu\mathrm{J}$ pulse energy, operating at $532\,\mathrm{nm}$. It is mounted next to the LIDAR mirror with a detector module in its focus. The assembly is mounted on an equatorial robotic mount that closely tracks the MAGIC observing direction during operation. By pointing at a $\sim5^{\circ}$ offset to the MAGIC pointing direction, it is ensured that the laser pulses do not disturb gamma-ray science observations. The laser shoots every 4~min at a rate of $250\,\mathrm{Hz}$. The backscattered light is collected by a 61 cm borosilicate glass mirror and focused on the detector module at a 150\,cm distance. The detector module comprises an aperture diaphragm and a lens that parallelizes the light to pass a $3\,\mathrm{nm}$ bandwidth interference filter.  Subsequently, a second lens is used to focus the light onto a hybrid photodetector (HPD). The HPD features a quantum efficiency (QE) of more than $50\,\%$ at the laser wavelength~\citep{saito_very_2009}. The amplified HPD waveforms are recorded at $500\,\mathrm{MSamples\,s^{-1}}$ at a 14-bit resolution by a dedicated computer located in a room below the LIDAR telescope protective dome. Due to the good charge separation of the HPD, it is possible to extract the single photoelectron charge directly from the recorded data~\citep{fruck_galactic_2015}. At distances $\gtrsim 7\,\mathrm{km}$, it is possible to count the individual photoelectrons of the signal. For closer distances, photon pileup occurs, and consequently, charge integration instead of photon counting becomes necessary. The charge integration is calibrated using the single photoelectron charge obtained from the far-field photon counting of the same recorded signal.
This LIDAR system is able to characterise the aerosol extinction of light of the used wavelength up to altitudes of $\sim20\,\mathrm{km\,a.s.l.}$ for low zenith angle observations. A range resolution of $\sim50\,\mathrm{m}$ is achieved in the lower troposphere and $\sim100\,\mathrm{m}$ in the upper regions.

The retrieved aerosol extinction profiles are converted into integrated aerosol transmissions as functions of height, which are then used to eventwise correct the gamma-ray event energy and the IACT instrument response function~\citep{schmuckermaier_correcting_2023}. The LIDAR system has been absolutely calibrated with a correlated accuracy better than 0.01 and an uncorrelated one of $\sim$0.015 vertical aerosol optical depths~\citep{Gaug:2022}.

\subsection{CTAO FRAM}

The FRAM (F/Photometric Robotic Telescope) measures the integral atmospheric extinction using as reference sources a large number of stars observed simultaneously in a wide field of view by a robotic telescope. A similar idea, albeit using individual stars, has been used in atmospheric sciences~\citep{PEREZRAMIREZ20082733,amt-8-3789-2015}; however the wide-field method, originally developed at the Pierre Auger Observatory \citep{2021JInst..16P6027A}, is more suitable for atmospheric monitoring at optical observatories thanks to its ability to capture temporal and spatial changes in atmospheric transparency. The FRAM concept has been adapted to the different requirements (field of view and mode of operation) of the CTAO~\citep{Janecek:2015rba}, and it is being further developed as an integral part of the atmospheric calibration program for the CTAO. 

During CTAO operations, FRAMs will provide measurements of aerosol optical depth (AOD) in the field of view of the Cherenkov telescopes to allow an interpolation between LIDAR measurements of aerosol profiles and the detection of changes in atmospheric conditions as explained in the first part of this section. Scans from zenith to the horizon will be carried out occasionally (either in set intervals or based on needs determined by real-time data analysis), because a fit of a photometric model to stars at a large span of airmasses can provide immediate absolute calibration of the entire chain that includes the lens, camera, and the chosen photometric method. 

Ahead of the CTAO construction, prototypes of FRAMs were deployed at the future CTAO sites, as parts of an extensive site characterisation campaign. Two FRAMs were deployed at the CTAO's southern site in Chile and one at its northern site, the ORM on the Canary Island of La Palma. During this campaign, the CTAO FRAMs carried out repeated precise measurements of the vertical AOD (VAOD) using the aforementioned scans from zenith to horizon. Apart from providing absolute calibration, these scans also allowed for the determination of the VAOD with an uncertainty of 0.006 (correlated) + 0.005 (uncorrelated) \citep{2021AJ....162....6E}, barring possible issues with the non-horizontal layering of aerosols. The optical system of a CTAO FRAM consists of a large-format CCD camera and a 135 mm photographic lens, covering a field of view of 15 × 15 degrees. Images are taken in the Johnson B filter with a peak of transmission at 420 nm, which is also used as the reference wavelength for the VAOD measurements.

\section{Results and Discussion}

\subsection{Atmospheric Aerosol Transmission Profiles Measured by the MAGIC LIDAR}
\label{sec:profiles}

The MAGIC LIDAR data analysis was described in detail in~\citet{fruck_characterizing_2022}, where statistical distributions of vertical aerosol optical depths were also presented, separately for the ground layer and clouds.

Gamma-ray-induced air showers normally die out before they reach the aerosol ground layer, unless strong Saharan dust intrusions (so-called “calima”) events happen, which, however, severely impair the observation conditions of an IACT. Under acceptable conditions, we may assume that all the observable Cherenkov light are equally extinct by the ground layer~\citep{dorner2009,schmuckermaier_correcting_2023,Pecimotika:2023}; its exact shape is of little importance, as long as its aerosol optical depth (AOD) is accurately determined.  The case of clouds is very different though: depending on the cloud's altitude and thickness, parts of the observable Cherenkov light are affected by extinction, whereas light emitted below the cloud is not~\citep{Sobczynska:2014,Sobczynska:2020,schmuckermaier_correcting_2023}. Because gamma-ray showers of higher energy penetrate deeper into the atmosphere than those generated by less energetic gamma rays, this leads to a strong energy dependency of the effect of clouds on the energy bias, resolution, and effective area~\citep{Pecimotika:2023}, in addition to the height dependency of the cloud. 

Unfortunately, clouds have escaped so far a fair standardisation of their shape and extension in simulations for IACTs~\citep{Pecimotika:2023}. They come with a large variety of thicknesses and vertical extinction profiles but have been treated, so far, as uniform layers of a constant extinction coefficient, normally of 1~km thickness (the so-called “binary clouds”~\citep{Pecimotika:2023}). The true clouds found above the CTAO-N site are different from those above in several aspects. 

Here, we re-analysed a subsample of the 16500~clouds contained in the 7 yr data sample of the MAGIC LIDAR~\citep{fruck_characterizing_2022} and investigated their distribution of shapes. From the original sample, only single clouds were selected with a vertical optical depth larger than 0.03 and LIDAR pointing elevation angle higher than 30$^\circ$. Clouds above both thresholds can be reconstructed with all the details of their vertical extent. More importantly, differences between their precise shape and the one simulated may start to affect systematic errors in the CTAO simulation and reconstruction chain beyond the 2\% permitted contribution to the systematic bias of the gamma-ray energy reconstruction. The remaining 3703 clouds were reconstructed from the elastic LIDAR signal using an iterative Klett algorithm~\citep{klett1981}, varying the LIDAR ratio as long as the integrated retrieved extinction matched the vertical optical depth (VOD) obtained from the comparison of the LIDAR signals from the purely molecular atmosphere below and above the cloud (see~\citet{fruck_characterizing_2022} for details). No effort to incorporate a height-dependent LIDAR ratio was made. Nevertheless, the distribution of such retrieved (vertically constant) LIDAR ratios was found to match expectations well~\citep{fruck_characterizing_2022}. 

\begin{figure}[H]
\includegraphics[width=0.485\textwidth]{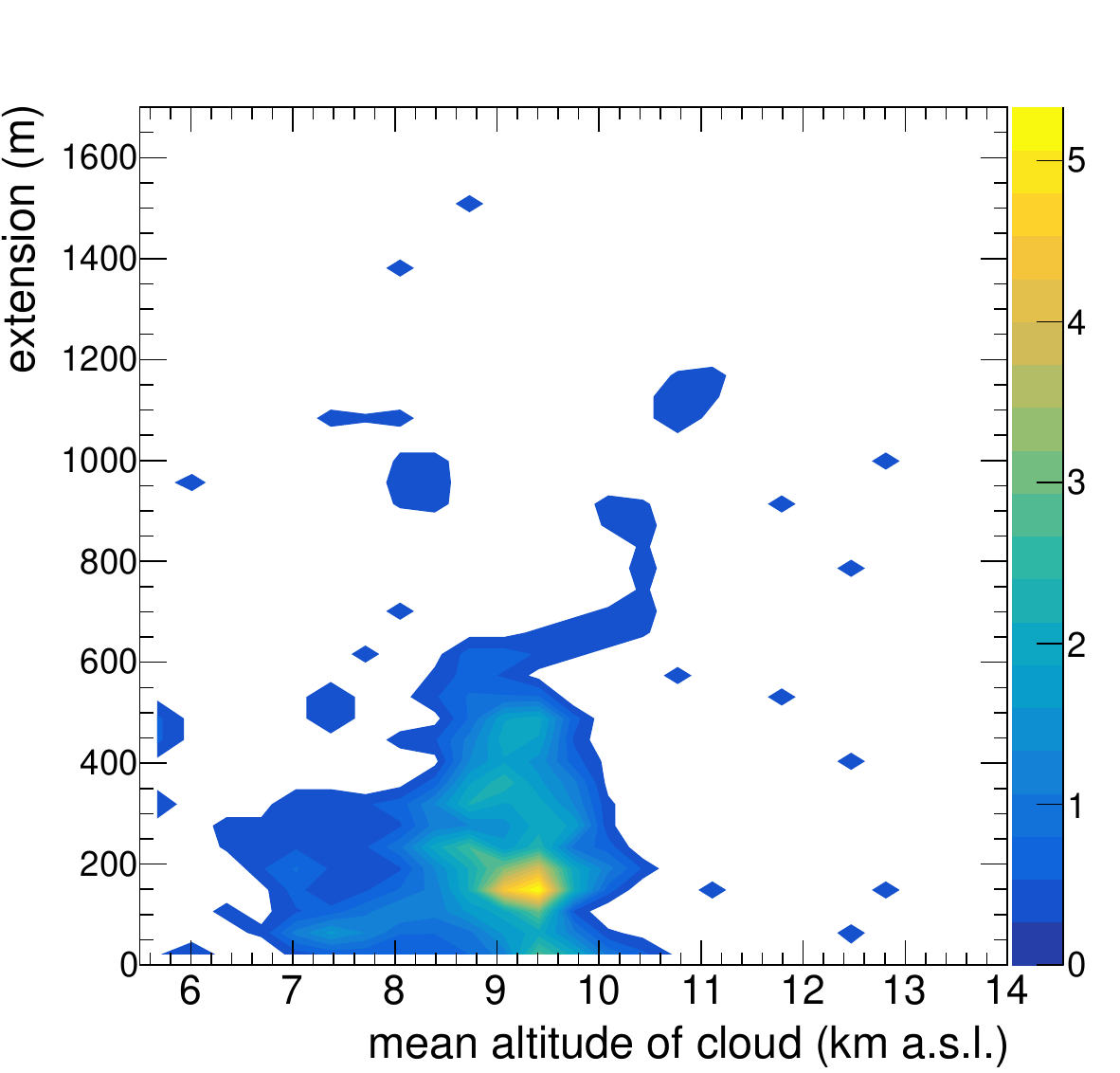}
\includegraphics[width=0.485\textwidth]{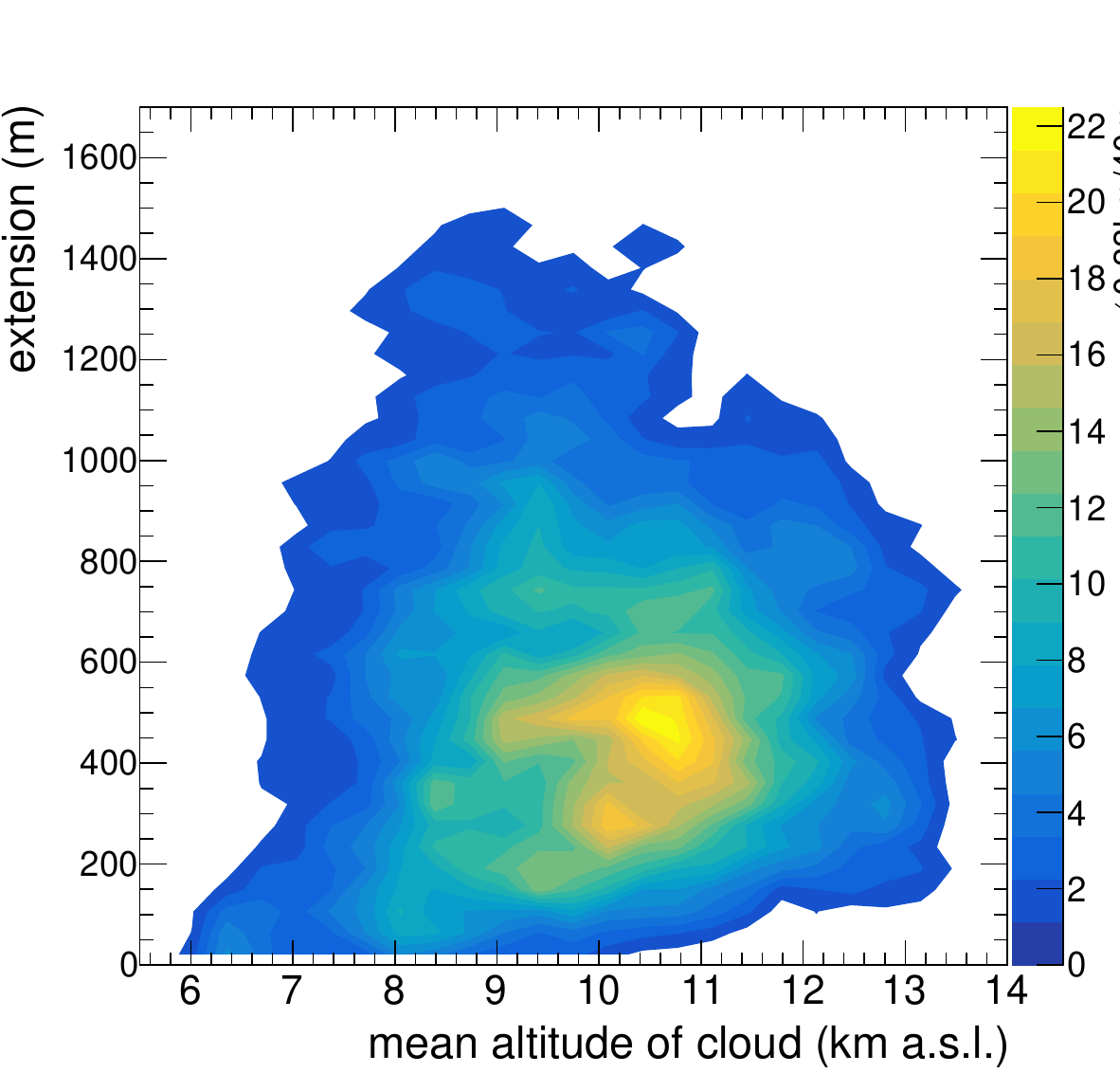}
\caption{\label{fig:cloudextension}Vertical standard deviation of clouds (“extension”) as a function of mean altitude. A smoothing on 3 × 3 consecutive cells (“k3a”) was applied. Left: summer clouds (July/August); right: clouds from the rest of the year.}
\end{figure}

With these caveats in mind, we first calculated the extinction profile of each cloud and from that, the mean cloud altitude $\overline{H}$ and the vertical standard deviation (hereafter “extension”), both obtained by using the extinction profile as weights. 
Figure~\ref{fig:cloudextension} shows the obtained cloud extensions as a function of their mean altitudes. One can immediately observe that during summer, when the Canary Islands enter the subtropical climate zone, the clouds are thinner (mostly $\lesssim$200~m in extension) and well located at altitudes between 7~km~a.s.l. up to 10~km~a.s.l., with a clear preference for $\sim$9.3~km~a.s.l. Clouds during the rest of the year show much larger variability. Starting from altitudes of about 8~km~a.s.l. to $\sim$11~km~a.s.l., a bimodal distribution is visible there, with a part of the clouds showing smaller extensions below 400~m, which increase as the clouds are found at higher altitudes, and another part with extensions of $\gtrsim$450~m, independent of altitude. We found no correlation between extension and LIDAR ratio, but a significant correlation of extensions vs. VODs. 

\begin{figure}[H]
\includegraphics[width=0.485\textwidth]{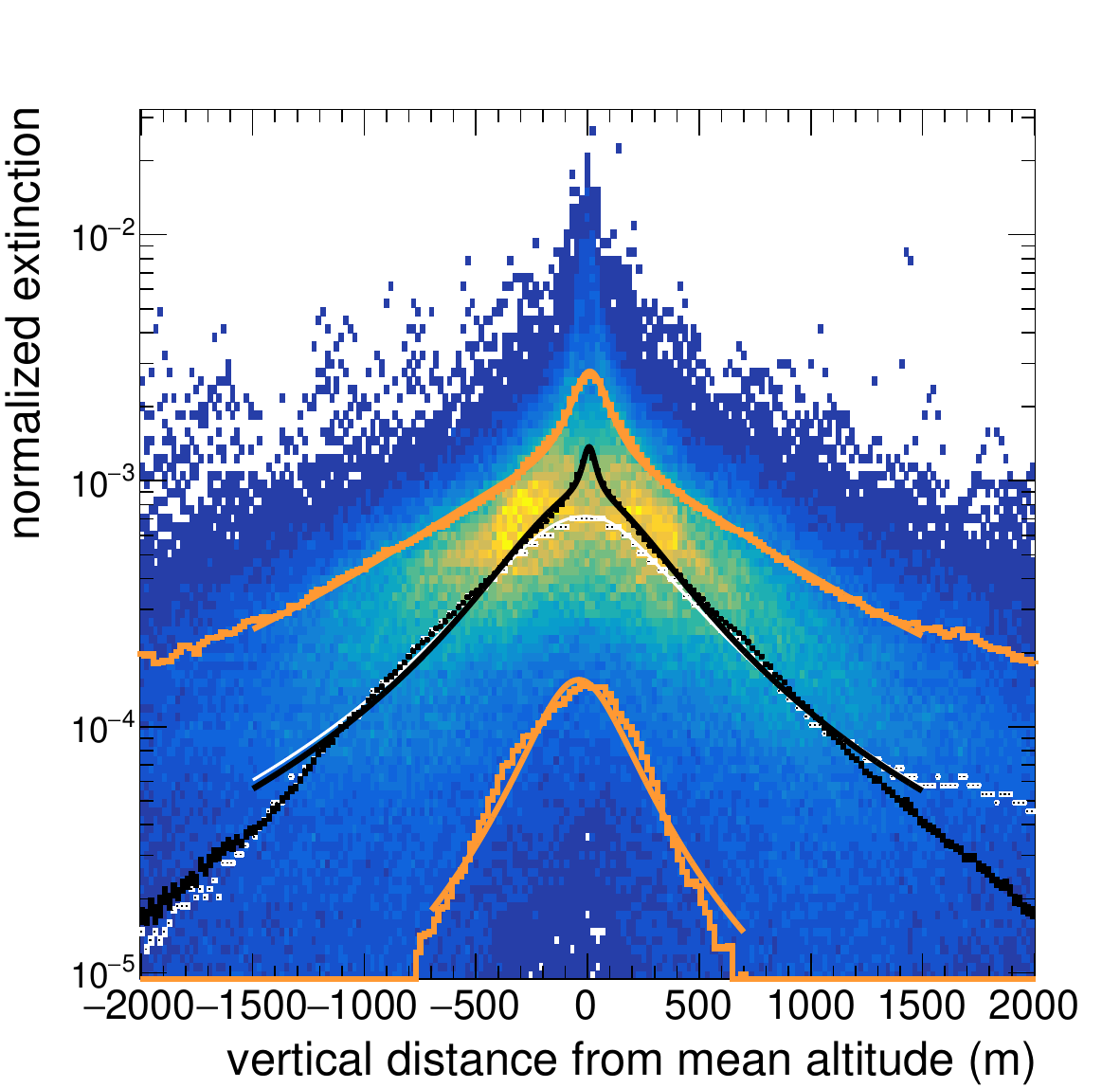}
\includegraphics[width=0.485\textwidth]{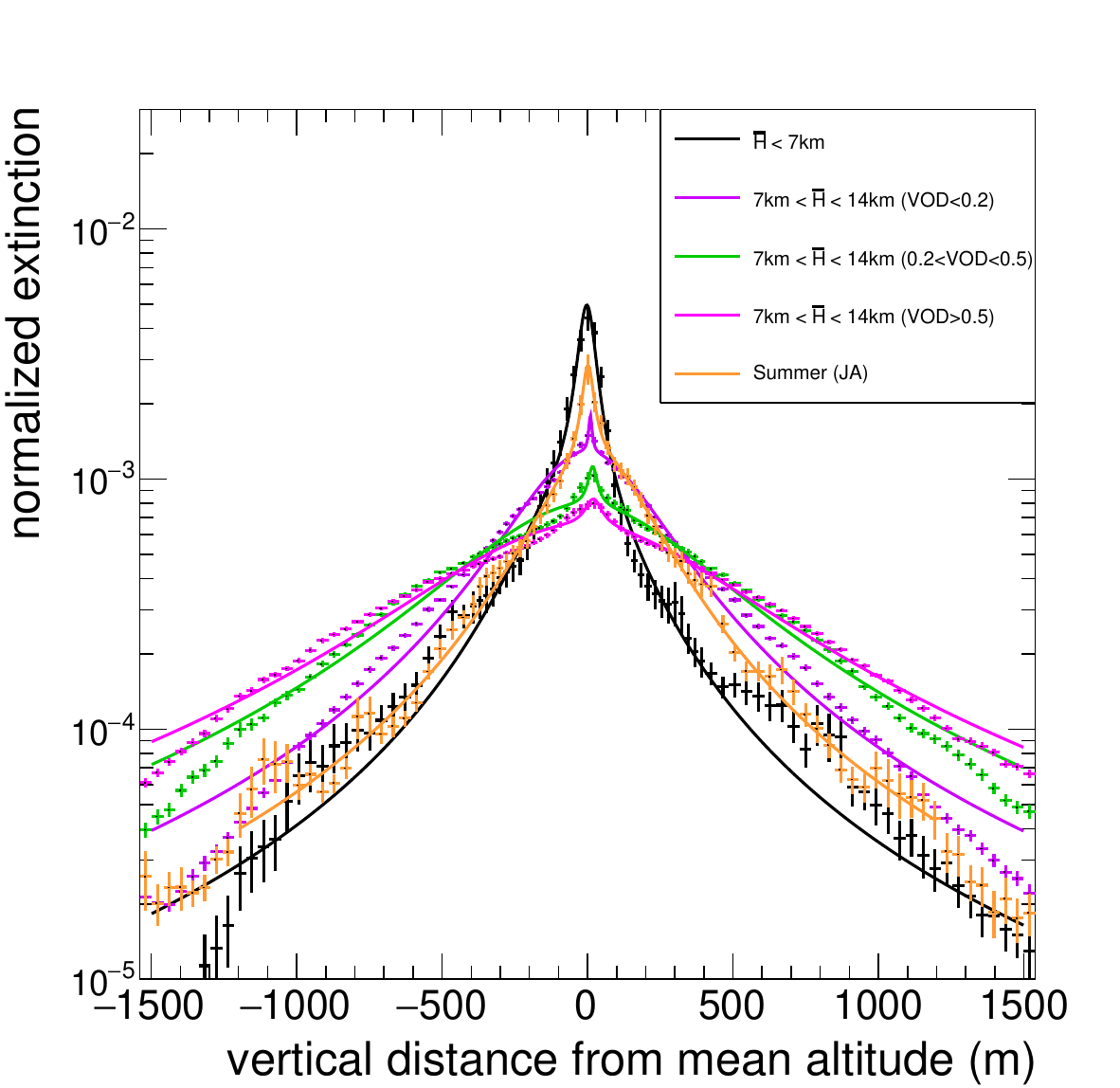}
\caption{\label{fig:cloudshapes}Distributions of normalised extinction vs. vertical distance from mean altitude (see text for definitions). Left: all extinction profiles (colour map), with the binwise means (black), medians (white), and 10\% and 90\% quantiles (orange). All parameters were fitted to Equation~(\ref{eq:doublelorentz}). Right: the mean extinction profiles for different cloud cases: low clouds ($\overline{H}<7~\mathrm{km~asl.}$, black), medium altitude clouds ($7~\mathrm{km~asl.}<\overline{H}<14~\mathrm{km~asl.}$) for three different VOD ranges: $\mathrm{VOD}<0.2$ (lilac), $0.2<\mathrm{VOD}<0.5$ (green) and $\mathrm{VOD}>0.5$ (pink), 
and summer clouds (orange). These profiles are also shown together with their fits to Equation~(\ref{eq:doublelorentz}). }
\end{figure}

In the following, we parameterized the shape and extensions of clouds, such that they could be used for dedicated simulation studies. To do so, we calculated for each cloud the extinction profile, normalised with respect to the cloud's VOD, and subtracted $\overline
{H}$ from the altitudes at which the extinction was obtained. The sum of all such “normalised extinction profiles” is shown in Figure~\ref{fig:cloudshapes} (left). At each vertical distance from the mean altitude, the statistical parameters mean, median, and the 10\% and 90\% quantiles were then calculated, taking into account that at higher distances from the mean altitude, many clouds had already reached their maximum extension, and the extinction needed to be counted correspondingly as zero. The four statistical parameters could be fitted reasonably well to a double Lorentzian distribution: 
\begin{equation}
f(x) = C \cdot \left( \dfrac{(1-A_2)\cdot \sigma_1 }{\pi \cdot  
 ((x-x_1)^2+\sigma_1^2)} + \dfrac{A_2\cdot \sigma_2 }{\pi \cdot  
 ((x-x_2)^2+\sigma_2^2)} \right) \quad. \label{eq:doublelorentz}
\end{equation}

Table~\ref{tab:globalfits} shows the parameters of Equation~(\ref{eq:doublelorentz}) fitted to the statistical parameters of Figure~\ref{fig:cloudshapes} (left). One can see that the mean and 95\% percentile required a double Lorentzian, whereas the median and 5\% percentile could be fitted with a single Lorentzian distribution. The 5\% quantile dropped off to negligible extinction values beyond $\pm$700~m from the mean cloud altitude. 

Finally, we divided the data set into four categories of clouds: low clouds ($\overline{H}<7~\mathrm{km~asl.}$), medium-altitude clouds ($7~\mathrm{km~asl.}<\overline{H}<14~\mathrm{km~asl.}$), and high clouds ($\overline{H}>14~\mathrm{km~asl.}$). The latter are not shown, because they barely influence the observed Cherenkov light~\citep{Sobczynska:2014}. The medium-altitude clouds were further divided into three VOD ranges, in order to take into account the observed dependence of cloud extension on the VOD. Finally, summer clouds were also treated separately. Figure~\ref{fig:cloudshapes} (right) shows the distribution of the mean of the normalised extinction for these cases, together with the results of a fit to Equation~(\ref{eq:doublelorentz}). Table~\ref{tab:globalfits2} lists the obtained fit results.
We observe that low clouds and summer clouds showed a non-negligible contribution of 47\% and 13\%, respectively, of thin clouds with Lorentzian spreads as low as a few tens of meters. The rest of the clouds were found with spreads between 200~m and 300~m, increasing to 400~m and $>$500~m as the VOD gets higher. 
It is interesting to observe that the vertical asymmetry of these thicker clouds was negative ($x_2 < 0$, i.e., the cloud appears skewed towards altitudes below $\overline{H}$) for all cases except for the summer clouds, where the skew was towards altitudes above the mean. 

\vspace{-13pt}

\vspace{-6pt}
\begin{table}[H] 
\caption{\label{tab:globalfits}
Fit results of normalised extinction profiles as a function of vertical distance from mean altitude, for all analysed clouds, fitted to Equation~(\ref{eq:doublelorentz}). }
\newcolumntype{C}{>{\centering\arraybackslash}X}
\begin{tabularx}{\textwidth}{lCCCCCC}
\toprule
\textbf{Case} & \boldmath$x_1$ & \boldmath$\sigma_1$ & \boldmath$A_2$ & \boldmath$x_2$ & \boldmath$\sigma_2$ & \textbf{C} \\  
    & \textbf{ (m)}   &\textbf{  (m)  }     & \textbf{ (1)}  & \textbf{ (m)   }&\textbf{  (m)    }   & \textbf{ (1) } \\
\midrule
Mean & 7 & 36 & 0.95 & $-$8 & 394 & 1.12  \\ 
Median & $-$26 & 448 & 0. & n.a. & n.a. & 1.00 \\
5\% Quantile &  $-$41 & 238 & 0. & n.a. & n.a. & 0.12  \\
95\% Quantile & 10 & 100. & 0.82 & $-$32 & 815 & 3.2 \\ 
\bottomrule
\end{tabularx}
\end{table}

\begin{table}[H] 
\caption{\label{tab:globalfits2}
Fit results of mean normalised extinction profiles as a function of vertical distance from mean altitude, for different subsamples of clouds, fitted to Equation~(\ref{eq:doublelorentz}). }
\newcolumntype{C}{>{\centering\arraybackslash}X}
\begin{tabularx}{\textwidth}{lCCCCCC}
\toprule
\textbf{Case} & \boldmath$x_1$ & \boldmath$\sigma_1$ & \boldmath$A_2$ & \boldmath$x_2$ & \boldmath$\sigma_2$ & \textbf{C} \\  
    & \textbf{ (m) }  & \textbf{ (m)}       & \textbf{ (1)  }&  \textbf{(m)}   &\textbf{  (m)  }     & \textbf{ (1) } \\
\midrule
{\small Low clouds}
& $-$2 & 35 & 0.53 & $-$46 & 210 & 0.98  \\ 
{\small Medium-altitude clouds,} 
{\small VOD $<$ 0.2} & 10 & 7 & 0.99 & $-$2 & 266 & 1.08 \\
{\small Medium-altitude clouds,}
{\small 0.2 $<$ VOD $<$ 0.5} & 18 & 19 & 0.98 & $-$12 & 473 & 1.18  \\
{\small Medium-altitude clouds,}
{\small VOD $>$ 0.5}  & 20 & 35 & 0.98 & $-$22 & 589 & 1.22 \\ 
{\small Summer clouds (July, August)} & 1 & 24 & 0.87 & 24 & 226 & 0.97 \\
\bottomrule
\end{tabularx}
\end{table}

\subsection{FRAM Aerosol Optical Depth Maps}

The ORM FRAM was installed in September 2018. Since early 2021, in addition to the precise VAOD measurements, it has also been operating in a mode that mimics the procedure planned for the CTAO, but in a simplified manner, as the details of the procedure are still being optimised. In this mode, the FRAM follows the field of view of the MAGIC telescopes and provides real-time AOD maps (presented as VAOD maps for practical reasons) for the operators of those telescopes. To facilitate this, the FRAM receives the pointing direction of the MAGIC LIDAR over the network. If this value has been updated within a set time period, it considers the LIDAR (and thus MAGIC) operational and continually takes single 30-second exposures in the Johnson B filter in this direction. Every 30 min, a scan from zenith to the horizon is taken and processed. In the standard VAOD analysis, the scans are processed retrospectively in batches of 50 to simultaneously extract some technical parameters of the telescope; in this real-time analysis, these parameters were fixed. The only free parameters in the fit of the photometric model for the observed stars were the aerosol extinction coefficient $k_\mathrm{A}$ (expressed in magnitudes, $k_\mathrm{A}=1.086\tau_\mathrm{A}$ where $\tau_\mathrm{A}$ is the VAOD) and zero point $Z$, which is the instrumental magnitude which would be measured for a star of magnitude zero outside of the atmosphere and which serves as the calibration constant. 

The stars in a scan were divided in bins in altitude above the horizon; if the RMS of residuals with respect to the fit in every bin was less than 0.2 magnitudes and the difference between the fitted zero point and the last known value was less than 1 magnitude, the zero point was adopted and used for the calculation of the AOD from the individual images until another scan passed the same criteria. These simple criteria were used as a rough filter to exclude scans affected by clouds, where the photometric model used to obtain the zero point did not hold, and the fit could produce arbitrary values depending on the position and thickness of the clouds.

\begin{figure}[H]
\includegraphics[width=0.97\textwidth]{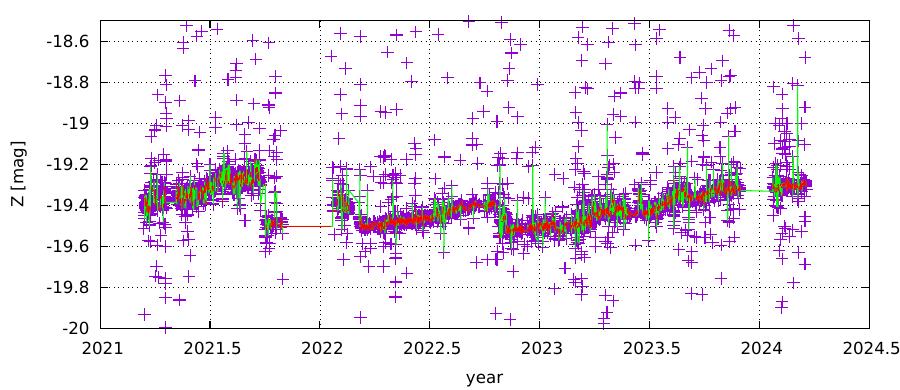}
\caption{Time series of zero points from scans. The green line connects values accepted during real-time processing; the red line connects values excluding outliers.\label{fig:z}}
\end{figure}

\begin{figure}[H]
\includegraphics[width=0.97\textwidth]{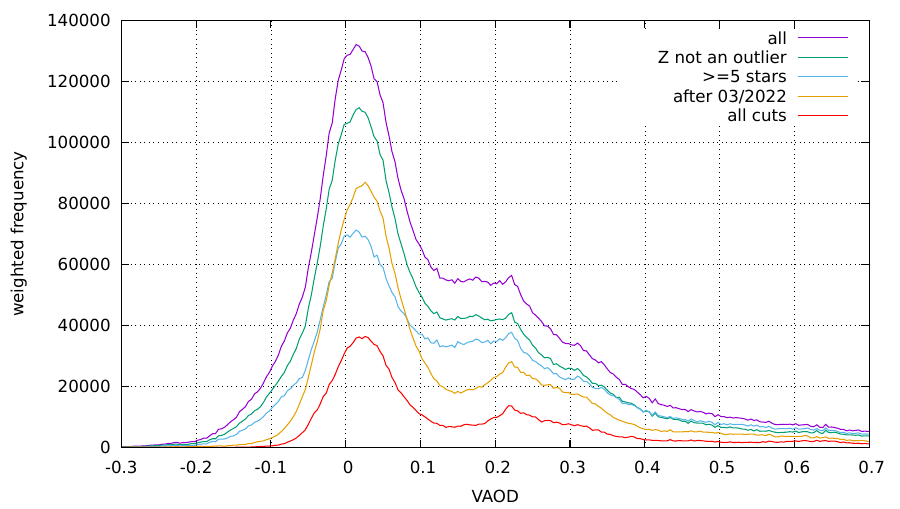}
\caption{Distribution of VAOD values measured in tiles with different cuts applied.\label{fig:hist1}}
\end{figure}

Figure~\ref{fig:z} shows the time series of zero points obtained from scans taken during the MAGIC FoV observations. The green line tracks the adopted zero-point values and thus connects all the scans that have been accepted. The time series shows many outliers, which can be defined as scans for which the resulting zero point is more than 0.02 magnitudes away from the average of the four nearest accepted zero points; the red line tracks zero points that would have been adopted if none of these outliers were accepted. The large number of outliers indicates that criteria have been set too loosely. The accumulated data will allow us to improve the procedure for future operation and select the optimal trade-off between the frequency and precision of a zero-point measurement. For the purpose of this analysis, we simply discarded the measurements where an outlier zero point was adapted, as this affected only about 20\% of the data (see Figure~\ref{fig:hist1}).

 The fast fluctuations are believed to occur mainly due to slight changes in the focus of the lens with the changing temperature through the night; the decrease in those from September 2021 was attributed to the change of the lens from a Zeiss 135/2 to a Samyang 135/2, which has far less optical aberrations, in particular off the centre of the image; a further change occurred in March 2022 due to an improved mechanical structure and optical alignment. The dependence of the zero point on focusing and alignment---in general on the point spread function (PSF) of stars in the image---can be significantly lowered using PSF-based photometry \citep{Negi_2022}, but this is still a work in progress, and it was not applied to real-time processing. The general rise in the zero point was mainly due to dirt accumulation on the lens, which reduced the sensitivity of the system. The October 2022 decrease corresponded to a site visit during which the lens was cleaned. The only other documented cleanings happened during the lens replacement and alignment described above; the reasons for the zero-point changes not always being monotonous in time are thus unclear.
 
\vspace{-6pt}

 The zero-point scans were taken even when the MAGIC telescope was not operated (and thus no FoV images were taken); the two long periods of downtime were due to the ash from the Tajogaite (formerly known as Cumbre Vieja) volcanic eruption, which contaminated the FRAM mechanics in 2021, and due to a cable corrosion issue at the end of 2023.

For a general star, the photometric model for the instrumental magnitude $m_\mathrm{inst}$ given the altitude and spectral sensitivity of the ORM FRAM in the B filter can be written as 
\begin{equation}
    m_\mathrm{inst}-B=Z+1.001k_\mathrm{A}A-0.02k_\mathrm{A}(B-V)A-0.0022k_\mathrm{A}^2A^2+\textrm{other terms} \label{eq:extinc}
\end{equation}
where A is the airmass, B and V are catalogue magnitudes for the given star, and the other terms are dependent either on the system properties or the molecular atmosphere and are assumed constant. In this approximation, we considered all aerosols to be contained in the ground layer and the \AA ngstr\"om coefficient, which describes the wavelength dependence of the VAOD, to be equal to one. The quadratic term is the first-order correction due to the finite bandwidth of the system; higher-order corrections are negligible. When the zero point is known from the previous accepted scan, the VAOD can be estimated from the measurement of a single star by solving the quadratic Equation~(\ref{eq:extinc}) and converting from magnitudes. For each image in the MAGIC FoV, we first divided the image based on the expected stars from the catalogue into tiles with a roughly equal number of stars using Voronoi tessellation \citep{2017EPJWC.14401012J} and then averaged the VAOD estimates over all stars within the tile which fulfil a set of quality cuts, thus creating a map that can be made available to MAGIC in real time.

\begin{figure}[H]
\includegraphics[width=0.98\textwidth]{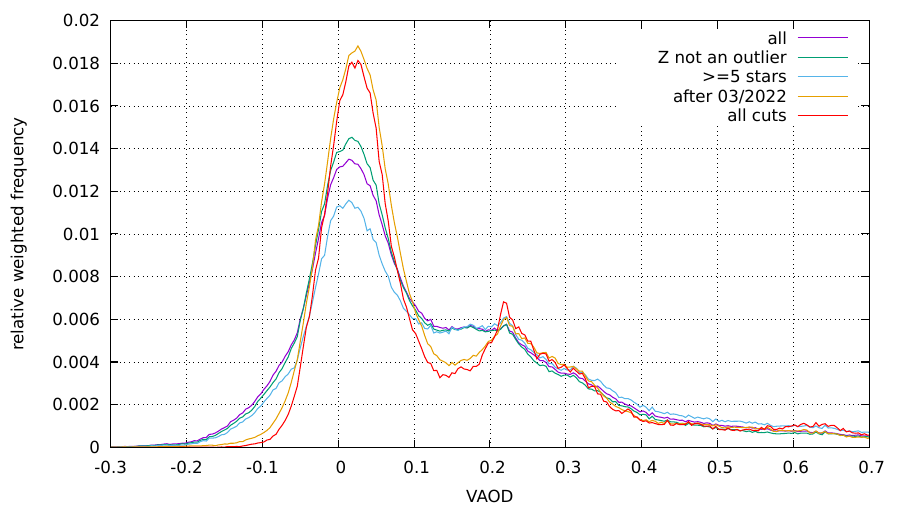}
\caption{Distribution  of VAOD values measured in tiles with different cuts applied, normalised to the number of entries after each cut.\label{fig:hist2}}
\end{figure}

During the three years of the program, VAOD was estimated for over 2.2 million tiles from over 67 thousand images. The distribution of the resulting VAOD values weighted by the number of stars in the tile is shown in Figure~\ref{fig:hist1}. This distribution was broad with many negative values, but this could be improved by imposing a series of quality cuts on the data. First, we required that the zero point used for the processing not be an outlier. Further, we specified that at least five stars should be used for the VAOD calculation in the tile---this was the goal of the Voronoi procedure in the first place, but due to a mistake in the processing pipeline, the cuts used to calculate the tile and to actually select the stars for processing were different. Finally, we selected data taken after 03/2022, that is, after the installation of the Samyang lens and clearing out the problems caused by the Tajogaite eruption, when the zero-point calibration was markedly less stable. The effects of these cuts become clearer in Figure~\ref{fig:hist2}, where the distributions are normalised with the number of entries after the cuts (nearly 300,000 entries, after all three cuts were applied). 

\begin{figure}[H]
\includegraphics[width=\textwidth]{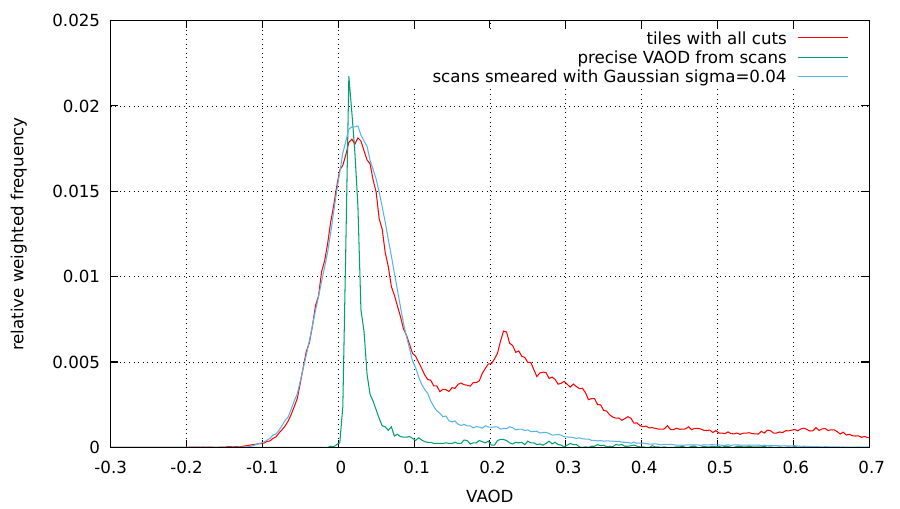}
\caption{Distribution of VAOD values measured in tiles after all cuts compared with the distribution of the precise VAOD measurements from scans without and with a smearing with a Gaussian with $\sigma=0.04$ \label{fig:hist3}}
\end{figure}

To assess the validity of the local VAOD measurements, we compared them with the precise VAOD data from scans taken over the same period (Figure~\ref{fig:hist3}). This distribution was much narrower, but we found that when smeared with a Gaussian function with $\sigma=0.04$, the shape of the peak was almost identical to that from the local measurements. The local measurements showed an excess for larger VAOD values, which is desirable---these are the clouds and other sources of excess extinction that the FoV-following measurements are specifically designed to uncover. We thus conclude that the local VAOD measurements reproduced well the precise VAOD measurements with a statistical fluctuation of 0.04 optical depth.

\begin{figure}[H]
\includegraphics[width=\textwidth]{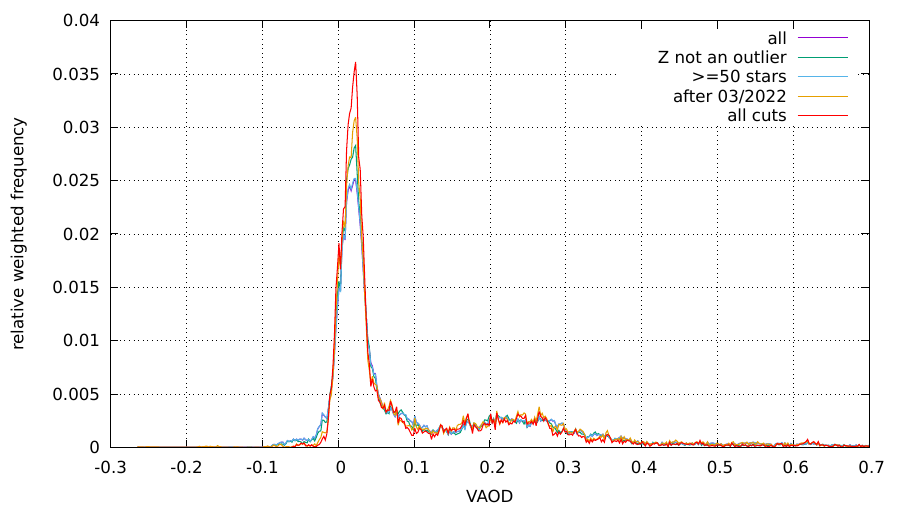}
\caption{Distribution of VAOD values measured in entire images with different cuts applied, normalised to the number of entire images after each cut. \label{fig:whist2}}
\end{figure}

\begin{figure}[H]
\includegraphics[width=\textwidth]{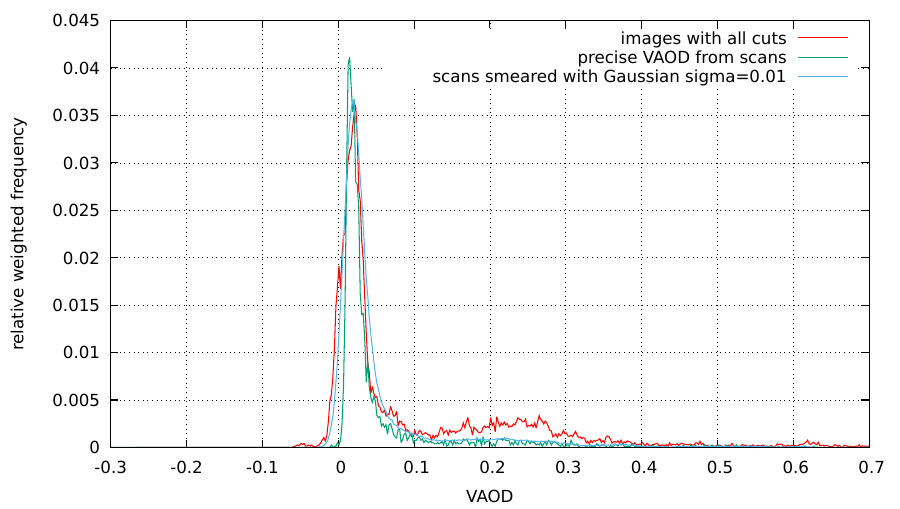}
\caption{Distribution of VAOD values measured in entire images after all cuts compared with the distribution of the precise VAOD measurements from scans without and with a smearing with a Gaussian with $\sigma=0.01$. \label{fig:whist3}}
\end{figure}

A large contribution to this fluctuation came from the rather small number of stars considered in each tile. This can be easily shown by plotting the distribution of VAOD values averaged over entire images and again imposing a similar series of cuts (Figure~\ref{fig:whist2}, already normalised by the number of passing images). As can be observed in Figure~\ref{fig:whist3}, the peak of the distribution could be reproduced by smearing the distribution of the precise VAOD, but this time, only with $\sigma=0.01$ optical depths. The secondary peak for whole images was lower and broader, showing that averaging over the images masked the presence of smaller clouds, and thus the tiling procedure was beneficial for a detailed understanding of the extinction.

Even though the fluctuations of 0.04 optical depth in the tiled measurements were relatively large, the found results were encouraging. We could almost immediately improve the result by improving the consistency between the construction of the tessellation and the star selection. Furthermore, we could optimise the selection cuts for the stars and the size of the tiles for an optimal trade-off between systematic errors, statistical precision, and angular resolution. Finally, this analysis was carried out using an older version of the processing pipeline based on the Tycho catalogue \citep{2000A&A...355L..27H}; the pipeline has already been modified to use GAIA \citep{2023A&A...674A...1G} as the source for star data, which is much more precise for fainter stars and thus allows the inclusion of stars up to 9.5 magnitudes, again increasing the available statistics for the tiles. 

\section{Conclusions and Future Perspectives}

From the presented results, we conclude that both LIDAR and FRAM are suitable instruments for the characterisation of the atmosphere above IACTs, especially important for the ORM, which is sometimes affected by fast changes in the presence, height, and structure of clouds and dust. In this work, a LIDAR with only one elastic channel was presented, currently used for the MAGIC, while in the future, a LIDAR with a full multiwavelength Raman LIDAR system will be used for the CTAO. In addition to better accuracy than in the case of an optical LIDAR, a Raman LIDAR will enable the determination of atmospheric temperature profiles that are relevant for studying climate changes.  

Taking into account that both MAGIC and CTAO-North are located at the same location at La Palma and are affected by the same atmospheric conditions, in the future a dedicated detailed study of correlations between the VAOD measurements taken by the MAGIC LIDAR and the CTAO FRAM would be interesting. Such a comparison would allow for a deeper understanding of the systematic errors of both instruments and pave the way to determining the accuracy with which atmospheric aerosol characterisation and monitoring can be achieved for the CTAO.

\vspace{6pt} 
%
%
\authorcontributions{Conceptualization, D.D.P., J.E. (Jan Ebr), and M.G.; methodology, D.D.P., J.E. (Jan Ebr), and M.G.; software, J.E. (Ji\v{r}\'{i} Eli\'{a}\v{s}ek), S.K., M.G., and A.B.; FRAM data: P.J., S.K., and J.E. (Ji\v{r}\'{i} Eli\'{a}\v{s}ek); formal analysis, J.E. (Jan Ebr), M.G., and M.K.; investigation, D.D.P. and M.K.; 
 data curation, M.G., J.E. (Jan Ebr), P.J., S.K., J.E. (Ji\v{r}\'{i} Eli\'{a}\v{s}ek), and M.K.; writing---original draft preparation, D.D.P., J.E. (Jan Ebr), M.G., and A.H.; writing---review and editing, all authors; visualization, M.G., J.E. (Jan Ebr), and A.B.; supervision, D.D.P.; project administration, D.D.P. and M.M.; funding acquisition, D.D.P., M.M., R.M., M.G., and J.E. (Jan Ebr). All authors have read and agreed to the published version of the manuscript.}

\funding{This research  was funded by the University of Rijeka Projects uniri-prirod-18-48 and uniri-iskusni-prirod-23-144, Croatian National Foundation (HRZZ) project IP-2022-10-4595, the Spanish grant PID2022-139117NB-C43, funded by MCIN/AEI/10.13039/501100011033/FEDER, UE, the German BMBF and MPG, the Departament de Recerca i Universitats de la Generalitat de Catalunya (grant SGR2021 00607), the infrastructure and research projects by Ministry of Education, Youth and Sports of the Czech Republic, and regional funds of the European Union, MEYS LM2023047 and EU/MEYS  CZ.02.01.01/00/22\_008/0004632 and CZ.02.01.01/00/22\_008/0004596.
}

\dataavailability{The data supporting the conclusions of this article will be made available by the authors on request.}

\acknowledgments{This work would have been impossible without the support of our colleagues from the MAGIC collaboration and the CTAO Consortium, which we gratefully acknowledge.
We would like to thank the Instituto de Astrof\'isica de Canarias for the excellent working conditions at the Observatorio del Roque de los Muchachos in La Palma. 
We also thank the funding agencies and institutions mentioned in the above section (Funding) for their financial support.}

\conflictsofinterest{The authors declare no conflicts of interest. The funders had no role in the design of the study; in the collection, analyses, or interpretation of data; in the writing of the manuscript; or in the decision to publish the results.}

\abbreviations{Abbreviations}{
The following abbreviations are used in this manuscript:\\

\noindent 
\begin{tabular}{@{}ll}
AGN& Active Galactic Nucleus\\
AOD& Aerosol Optical Depth\\
CCD& Charge Coupled Device\\
CTAO& Cherenkov Telescope Array Observatory\\
FoV& Field of view\\
FRAM& F/Photometric Robotic Telescope\\
GRB& Gamma-ray burst\\
HPD& Hybrid photodetector\\
IACT& Imaging Atmospheric Cherenkov Telescope\\
IRF& Instrument response function\\
LIDAR& LIght Detection And Ranging\\
LIV& Lorenz Invariance Violation\\
MAGIC& Major Atmospheric Gamma Imaging Cherenkov\\
MC& Monte Carlo\\
MWL& Multiwavelength\\
OB& Observation block\\
ORM& Observatorio del Roque de los Muchachos\\
PSF& Point spread function\\
QE& Quantum efficiency\\
STI& Stable Time Interval\\
ToO& Target of Opportunity\\
VAOD& Vertical aerosol optical depth\\
VHE& Very High Energy\end{tabular}
}


\begin{adjustwidth}{-\extralength}{0cm}

\reftitle{References}


\PublishersNote{}
\end{adjustwidth}
\end{document}